\documentclass[aps,showpacs,preprint]{revtex4}

\begin{document}

\title{Peculiarities of the Canonical Analysis of the First Order Form of 
the Einstein-Hilbert Action in Two Dimensions in Terms of\\ 
the  Metric Tensor or the Metric Density}

\author{N. Kiriushcheva}
\email{nkiriush@uwo.ca}
\affiliation{Department of Mathematics,}

\author{S.V. Kuzmin}
\email{skuzmin@uwo.ca}

\author{D.G.C. McKeon}
\email{dgmckeo2@uwo.ca}
\affiliation{Department of Applied Mathematics, 
University of Western Ontario, London, N6A~5B7 Canada}

\date{\today}

\begin{abstract}
The peculiarities of doing a canonical analysis of the 
first order formulation of the Einstein-Hilbert action in terms of either the 
metric tensor $g^{\alpha \beta}$ or the metric density $h^{\alpha \beta}=
\sqrt{-g}g^{\alpha \beta}$ along with the affine connection are discussed. 
It is shown that the difference between using $g^{\alpha \beta}$ as opposed to
$h^{\alpha \beta}$ appears only in two spacetime dimensions. Despite there 
being a different number of constraints in these two approaches, both
formulations result in there being a local Poisson brackets algebra of 
constraints with field independent structure
constants, closed off shell generators of gauge transformations and off shell
invariance of the action. The formulation in terms of the metric tensor 
is analyzed in detail and compared with earlier results obtained using the 
metric density. 
The gauge transformations, obtained from the full set of first class 
constraints, are different from a diffeomorphism transformation in both cases.

\end{abstract}

\maketitle

The Einstein-Hilbert (EH) action in $d$ spacetime dimensions

\begin{equation}
\label{soEH}S_d\left(g^{\alpha \beta}\right)=\int d^dx\sqrt{-g}R 
\end{equation}

is known to have very interesting though complicated structure. In two
dimensions ($2D$) the action (\ref{soEH}) is a total divergence \cite{LL} when
$R$ is expressed solely in terms of the metric. This does not allow for a 
study of the basic questions of quantum gravity in the $2D$ limit of 
(\ref{soEH}). Therefore, a variety of alternate $2D$ models of gravity have 
been developed and investigated. (For a recent review see \cite{GKV2002}.) 
Such $2D$ models have an action which is not just a total divergence. 
In higher dimensions the EH action of (\ref{soEH}) is the most interesting
model. The canonical treatment of (\ref{soEH}) is not only technically 
complicated, but is peculiar because of the presence of terms
with second order derivatives which are not encounted in ordinary gauge 
theories. In the action of (\ref {soEH}), such terms can be put in the form of 
a total divergence without any additional first order terms \cite{LL} but they 
are needed to retain invariance of the action under
general coordinate transformations and must be treated carefully.
(For discussion of this problem see \cite{RT}.) 

An efficient and well-known way to avoid this difficulty is to use a first
order formulation of (\ref{soEH}). In this approach the Dirac procedure 
\cite{Dirac1964} can be applied without modification, as in first order 
formulations of ordinary gauge theories \cite{Sund}. The first order 
formulation of (\ref{soEH}) was proposed by Einstein \cite{Einstein1925} 
(through this is often attributed to Palatini \cite{Pal}). In the formulation 
of \cite{Einstein1925} the Lagrange density takes a simple form

\begin{equation}
\label{foEH}L_d=\sqrt{-g}R=h^{\alpha \beta }\left( \Gamma _{\alpha \beta
,\lambda }^\lambda -\Gamma _{\alpha \lambda ,\beta }^\lambda +\Gamma
_{\sigma \lambda }^\lambda \Gamma _{\alpha \beta }^\sigma -\Gamma _{\sigma
\alpha }^\lambda \Gamma _{\lambda \beta }^\sigma \right), 
\end{equation}

where the symmetric affine connection $\Gamma _{\alpha \beta}^\lambda$ is 
considered as an independent field without identifying it with the Christoffel 
symbol, and $h^{\alpha \beta}=\sqrt{-g}g^{\alpha \beta }$ is the metric 
density. This formulation was used for the first time by Arnowitt, Deser and 
Misner (ADM) \cite{ADM1,ADM2} to study the canonical properties of $S_4$.

In more than two dimensions the second order (\ref{soEH}) and first order 
(\ref{foEH}) formulations of EH action are equivalent \cite{Einstein1925}. 
In $2D$ they are different; the equation of motion for 
$\Gamma_{\alpha \beta}^{\lambda}$ in $2D$ no longer implies that it equals the 
Christoffel symbol. This was analyzed using the Lagrangian approach in 
\cite{Mann}. It is interesting to perform a canonical analysis of the $2D$ EH 
action in first order form; this is possible as it is no longer a total 
divergence. In any case, (\ref{foEH}) is formally valid in all dimensions, 
even though its properties can be quite different in
special dimensions. Moreover, the lowest dimensional version of (\ref{foEH})
is a $2D$ limit of an action which is equivalent in all higher dimensions to
the second order formulation and we would expect that some features would be 
common to the first order formulation in all dimensions. The simplicity 
of (\ref{foEH}) in $2D$ (e.g., the number of fields is even less than in the 
first order formulation of the $4D$ Maxwell Lagrangian \cite{Sund})
allows for a straightforward application of the Dirac procedure without making
any {\it a priori} assumptions or restrictions. The canonical analysis 
of $L_2\left(h,\Gamma\right)$ in (\ref{foEH}) using the Dirac procedure, 
treating the metric density as an independent field, was performed in 
\cite{KKM1}. The resulting structure is similar to what is encounted in 
ordinary gauge theories: there is a local algebra of constraints with field 
independent structure constants, a closed off shell algebra of generators, 
and a gauge transformation
that preserves the exact invariance of $L_2$. The gauge transformation 
implied by the first class constraints is different from a general coordinate 
transformation. 

A local algebra of constraints was obtained earlier in dilaton gravity 
\cite{GKV2002} but, unlike the algebra of \cite{KKM1}, this algebra has field 
dependent structure constants.

In \cite{KKM1} (as in \cite{Einstein1925},\cite{ADM1}) $\Gamma _{\alpha
\beta }^\sigma $ and the metric density $h^{\alpha \beta }$ were used as
an independent set of variables. Is it possible that all the desirable 
canonical properties obtained in \cite{KKM1} are 
just an artifact of using $h^{\alpha \beta }$ instead of $g^{\alpha \beta }$? 
This needs to be investigated, as the functional Jacobian 
$\frac{\delta h^{\alpha \beta}}{\delta g^{\mu \nu}}$, 
corresponding to a change of variables $h^{\alpha \beta}=\sqrt{-g}
g^{\alpha \beta}$, is field dependent in $d>2$ and is singular in $2D$, 
since in $2D$ the components of $h^{\alpha \beta}$ are not independent, being 
restricted by the condition $\det \left(h^{\alpha \beta}\right)=-1$.    
The main goal of this letter is to trace the difference in the
canonical analysis when using  $g^{\alpha \beta}$ in place of 
$h^{\alpha \beta}$ and to demonstrate that, despite there being
of quite a different constraint structure when using $g^{\alpha \beta}$, 
all canonical properties found in \cite {KKM1} remain intact and they are 
not just an artifact of using $h^{\alpha \beta}$.

Straightforward application of the Dirac procedure in \cite{KKM1}
automatically gave a local algebra of constraints with field independent
structure constants. To preserve these properties and to also have off shell 
closure of the algebra of generators and exact invariance of the Lagrangian, 
it was necessary to choose a simple linear transformation of the affine 
connections as dynamical variables. These transformations were found 
in component form in \cite{KKM1} but they can be recast into covariant form:

\begin{equation}
\label{CT2}\xi_{\alpha \beta}^\lambda = \Gamma _{\alpha \beta}^\lambda 
-\frac{1}{2}\left(\delta_\alpha^\lambda \Gamma _{\beta \sigma}^\sigma + 
\delta_\beta^\lambda \Gamma_{\alpha \sigma}^\sigma \right),
\end{equation}

so that
\begin{equation}
\label{ICT2}\Gamma_{\alpha \beta}^\lambda = \xi _{\alpha \beta}^\lambda 
-\frac{1}{d-1}\left(\delta_\alpha^\lambda \xi _{\beta \sigma}^\sigma + 
\delta_\beta^\lambda \xi_{\alpha \sigma}^\sigma \right).
\end{equation}

Upon substitution into (\ref{foEH}) we obtain

\begin{equation}
\label{LEH}\tilde L_d\left(g,\xi\right) = h^{\alpha \beta}\left( 
\xi _{\alpha \beta,\lambda }^\lambda -
\xi _{\alpha \sigma}^\lambda \xi_{\beta \lambda }^\sigma + 
\frac{1}{d-1}\xi _{\alpha \lambda }^\lambda \xi_{\beta 
\sigma }^\sigma \right). 
\end{equation}

Equation (\ref{LEH}) provides an alternative first order formulation of the EH 
action, as upon substitution of the solution of 
$\frac{\delta \tilde L_d}{\delta \xi_{\alpha \beta}^\lambda}=0$
into $\frac{\delta \tilde L_d}{\delta g^{\alpha \beta}}=0$ 
one can obtain the usual Einstein field equations without 
any references to $\Gamma_{\alpha \beta}^\lambda$; this 
calculation is actually simplier than when done using 
$\Gamma_{\alpha \beta}^\lambda$.
The main advantage of (\ref{LEH}) is the ``diagonal'' form of the derivative 
part of the EH action. This is especially well suited for a canonical analysis 
as the equation

\begin{equation}
\label{h}h^{\alpha \beta} \xi _{\alpha \beta,\lambda }^\lambda =
h^{\alpha \beta} \dot \xi _{\alpha \beta}^0 + h^{\alpha \beta} \xi _{\alpha 
\beta, k }^k 
\end{equation}

shows that there is a simple separation of the components of 
$\xi _{\alpha \beta}^\lambda$ 
into those which are dynamical ( $\xi _{\alpha \beta}^0$) and those which are
non-dynamical ($\xi _{\alpha \beta}^k$). (Latin indices indicate  
spatial components.) When using the variables  $h$ and $\Gamma$, the 
decomposition 
(\ref{h}) is not as simple, since some components of $\Gamma$ enter $L_d$ with 
both spatial and temporal derivatives, making a straightforward Dirac 
analysis more difficult.

A full Dirac analysis of $\tilde L_d$ for $d>2$ is beyond the scope of
this letter; we shall discuss only the first stage of the Dirac procedure 
and the question of whether a formulation using the metric tensor is 
equivalent to one using the metric tensor density. First introducing momenta 
conjugate to all fields

\begin{equation}
\label{Md}\pi_{\alpha \beta}\left( g^{\alpha \beta}\right), 
\Pi_{0}^{\alpha \beta}\left( \xi_{\alpha \beta}^{0}\right), 
\Pi_{k}^{\alpha \beta}\left( \xi_{\alpha \beta}^{k}\right)
\end{equation}

and using (\ref{h}) we immediately obtain the $\frac{1}{2}d\left(d+1\right)^2$
primary constraints

\begin{equation}
\label{Pd}\pi _{\alpha \beta} \approx 0,\Pi_k^{\alpha \beta} \approx 0,
\Pi_0^{\alpha \beta} - \sqrt{-g} g^{\alpha \beta} \approx 0
\end{equation}

If the $d\left(d+1\right)$ by $d\left(d+1\right)$  matrix
 
\begin{equation}
\label{1}\tilde M_d =\left( \left\{\phi,\tilde \phi \right\} \right) 
\end{equation}

built from the non-zero Poisson brackets (PB) among the primary constraints 
$\phi, \tilde \phi \in \left(\pi _{\alpha \beta},\Pi_0^{\alpha \beta}
 - \sqrt{-g} g^{\alpha \beta} \right)$ is invertible, these constraints are 
all second class. If the rank of the matrix (\ref{1}) is $r$, then 
there are $d\left(d+1\right)-r$ first class constraints. Moreover, 
all these constraints are constraints of a special form in which one
constraint is $\pi_{\alpha \beta}\approx 0$ and the other has 
$g^{\alpha \beta }$ as a function of the other dynamical variables. (See 
\cite{Dirac1964} and for a more detailed and general discussion 
\cite{GT1990}.) 
For such constraints, if $\det \tilde M_d\neq 0$ we can eliminate the momenta 
$\pi_{\alpha \beta}$ by setting them equal to zero and then solving for 
$g^{\alpha \beta }=g^{\alpha \beta }\left( \Pi_0^{\gamma \sigma}\right) $
and subsequently substituting this expression into the Hamiltonian and all 
the remaining 
constraints. (This is the so-called Dirac or Hamiltonian reduction in its 
simplest form since in this case the Dirac brackets are equivalent to PB for 
the remaining variables.) For 
$\tilde L_d$ it is not even necessary to solve any equation for 
$g^{\alpha \beta }$ as it enters the Hamiltonian in the combination 
$\sqrt {-g}g^{\alpha \beta}$ which is what is present 
in the second class  primary constraints  and the solutions for such 
combinations are given immediately. Once the canonical analysis gives the
gauge transformation for  $\Pi_{0}^{\alpha \beta}$, the
equality $\Pi_{0}^{\alpha \beta}=\sqrt{-g} g^{\alpha \beta}$ shows  
how $g^{\alpha \beta}$ itself transforms under a gauge transformation. 

In $2D$ (and only in $2D$) the matrix (\ref{1}) is singular when the metric is 
used as an independent field and consequently in $2D$, formulations in terms 
of $g^{\alpha \beta }$ and $h^{\alpha \beta }$ are distinct; we cannot 
determine the gauge transformation of $g^{\alpha \beta}$ from the gauge 
transformation of $h^{\alpha \beta}$ found in \cite{KKM1}. In $2D$ the rank 
of the $6\times 6$ matrix $\tilde M_2$ is four and thus only two pairs of 
constraints constitute a second class subset of constraints that have
a special form that allows two variables to be eliminated by using Dirac
reduction. (When the $2D$ EH action is formulated in terms of 
$h^{\alpha \beta}$, the rank of $\tilde M_{2}$ is six and hence there are 
three such pairs of second class constraints \cite {KKM1}.)

The Lagrangian density when written in component form is given by
$$
\tilde L_2\left(g,\xi\right)=h^{11}\dot \xi_{11}^0 + 2 h^{01}\dot 
\xi _{01}^0 + h^{00}\dot \xi_{00}^0 - H
$$
where 
$$
H = \xi_{11}^1\left( h_{,1}^{11} - 2 h^{11}\xi_{01}^0 - 2 h^{01} 
\xi_{00}^0 \right)
$$
\begin{equation}
\label{3}+2\xi_{01}^1\left( h_{,1}^{01} + h^{11} \xi_{11}^0 - 
h^{00} \xi_{00}^0\right) +\xi _{00}^1 \left( h_{,1}^{00} + 2 h^{01} 
\xi_{11}^0 + 2 h^{00}\xi_{01}^0 \right). 
\end{equation}

Note, that in this expression $h^{\alpha \beta }$ is just a short form for 
$\sqrt{-g}g^{\alpha \beta }$ (and it is not treated as an independent 
variable) and 
integration by parts has been performed in the spatial derivatives. Introducing
the momenta (\ref{Md}) conjugate to all fields, we obtain nine primary 
constraints

\begin{equation}
\label{4}\pi_{00}\approx 0, \pi_{01}\approx 0, \pi_{11}\approx 0, 
\end{equation}

\begin{equation}
\label{5}\Pi_1^{11} \approx 0,\Pi _1^{01}\approx 0,\Pi_1^{00} \approx 0, 
\end{equation}

\begin{equation}
\label{6}\Pi_0^{11}-\sqrt{-g}g^{11} \approx 0,\Pi_0^{01}-\sqrt{-g}g^{01} 
\approx 0,\Pi_0^{00} -\sqrt{-g}g^{00} \approx 0. 
\end{equation}

We employ the standard fundamental PB 

\begin{equation}
\label{7}\left\{ g^{\alpha \beta },\pi _{\mu \nu }\right\} = \Delta _{\mu \nu
}^{\alpha \beta }, \left\{ \Gamma_{\alpha \beta }^\lambda ,\Pi _\sigma^
{\mu \nu }\right\} = \delta_\sigma^\lambda \Delta ^{\mu \nu}_{\alpha \beta }, 
\end{equation}

where $\Delta _{\mu \nu}^{\alpha \beta} = \frac{1}{2} \left(\delta_\mu^\alpha 
\delta_\nu^\beta + \delta_\mu^\beta \delta_\nu^\alpha \right)$.  

The matrix of PB among the primary constraints (\ref{4},\ref{6}) has rank four 
and so two pairs of constraints constitute a second class subset. 
We pick the following pairs

\begin{equation}
\label{8}\pi _{00}\approx 0,\Pi_0^{00} -\sqrt{-g}g^{00}\approx 0; 
\pi_{01}\approx 0,\Pi_0^{01} - \sqrt{-g}g^{01}\approx 0. 
\end{equation}

From these we can eliminate $\pi_{00}$, $g^{00}$, $\pi^{01}$ and $g^{01}$ 
using the strong equations \cite{Dirac1964}

\begin{equation}
\label{9}\pi _{00}=0,g^{00}=g^{11}\frac{\Pi_0^{00} \Pi_0^{00} }
{\Pi_0 ^{01} \Pi_0 ^{01}-1}; \pi_{01}=0,g^{01}=g^{11}\frac{\Pi_0^{00} 
\Pi_0 ^{01}}{\Pi_0 ^{01} \Pi_0^{01}-1}. 
\end{equation}

To eliminate $g^{01}$ and $g^{00}$ in the Hamiltonian, we actually need only 
the combinations 

\begin{equation}
\label{10}h^{01}\equiv \sqrt{-g}g^{01}=\Pi_0 ^{01},h^{00}\equiv \sqrt{-g}%
g^{00}=\Pi_0^{00} , 
\end{equation}

For the first constraint of (\ref{6}) and the Hamiltonian we also need
the expression for $\sqrt{-g}g^{11}$, which upon using (\ref{9}) equals 

\begin{equation}
\label{12}h^{11}\equiv \sqrt{-g}g^{11}=\frac{1}{\Pi_0^{00}} \left( \Pi_0 ^{01} 
\Pi_0^{01}-1 \right). 
\end{equation}

This also follows from equality 

\begin{equation}
\label{11}h^{00}h^{11}-h^{01}h^{01}=-\det \left( g_{\alpha \beta }\right)
\left(g^{00}g^{11}-g^{01}g^{01}\right) =-1. 
\end{equation}

After the first stage of the Dirac reduction (substitution of (\ref{9}) into 
both the constraints and the Hamiltonian), we are left with five primary 
constraints and we have eliminated two canonical pairs of variables 
(corresponding to the four primary second class constraints). These five 
constraints are

\begin{equation}
\label{13}\Pi_k^{\alpha \beta} \approx 0,\pi _{11}\approx 0, 
\Pi_0^{11}-\frac {1}{\Pi_0^{00}}\left( \Pi_0^{01}\Pi_0^{01}-1\right)\approx 0, 
\end{equation}

and the Hamiltonian is

\begin{equation}
\label{14}H_c=-\xi_{11}^1 \chi_1^{11}-2\xi_{01}^1 \chi_1^{01}
-\xi_{00}^1 \chi_1^{00}  
\end{equation}

where
$$
\chi_1^{11}=-\left(\left( \frac {1}{\Pi_0^{00}} \left( \Pi_0^{01} \Pi_0^{01}
-1\right)\right) _{,1}- \frac {2}{\Pi_0^{00}} \left( \Pi_0^{01}
\Pi_0^{01}-1 \right) \xi_{01}^0 - 2 \Pi_0^{01}\xi_{00}^0\right), 
$$
$$
\chi_1^{01}=-\left(\Pi _{0,1}^{01}+\frac {1}{\Pi_0^{00}} \left( \Pi_0^{01}
\Pi_0^{01}-1\right) \xi _{11}^0-\Pi_0^{00} \xi_{00}^0 \right), 
$$
$$
\chi_1^{00}=-\left( \Pi _{0,1}^{00}+2\Pi_0^{00} \xi_{01}^0+2\Pi_0^{01}
\xi _{11}^0\right). 
$$

Note, that the canonical pair $(g^{11},\pi_{11})$ is not explicitly present 
in $H_c$ even though it was not eliminated by the process of Dirac reduction. 
The role of this ``hidden variable'' becomes apparent later when the gauge 
transformation of $g^{\alpha \beta}$ is calculated. Also, if we include a
cosmological term linear in $\sqrt{-g}$ in $H_c$, substitution of (\ref{9}) 
gives $\sqrt{-g}=\frac {\Pi_0^{01}\Pi_0^{01}-1}{g^{11}\Pi_0^{00}}$ and 
$H_c$ would then depend explicitly on $g^{11}$ affecting the 
constraint structure of the model.

All the primary constraints of (\ref {13}) have vanishing  PB among themselves 
so we go to the next step of the Dirac procedure and consider the persistence 
of the primary constraints in time. Three of the five primary first class 
constraints have a non-zero PB with the Hamiltonian, thus producing the 
secondary constraints

\begin{equation}
\label{15}\dot \Pi_1^{11}=\left\{ \Pi_1^{11},H\right\} = \chi_1^{11}, 
\dot \Pi_1^{01} =\left\{ \Pi_1^{01} ,H\right\} = \chi_1^{01}, 
\dot \Pi _1^{00} = \left\{ \Pi _1^{00},H\right\} = \chi _1^{00} 
\end{equation}

The constraints (\ref{15}) have the same algebraic structure as the secondary 
constraints arising in the $h^{\alpha \beta}$ formulation \cite{KKM1}, that 
is, there is a local algebra of PB with field independent structure constants

\begin{equation}
\label{18}\left\{ \chi_1^{01} ,\chi _1^{00}\right\} =\chi _1^{00},
\left\{ \chi_1^{01} ,\chi_1^{11}\right\} =-\chi_1^{11},
\left\{ \chi_1^{11},\chi _1^{00}\right\} =2\chi_1^{01}. 
\end{equation}

The Hamiltonian of (\ref {14}) is a linear combination of these secondary
constraints and there are thus no tertiary constraints. (This is not the case 
when $d>2$.) The only non zero PB are those given by (\ref{18}) since all 
secondary constraints have a vanishing PB 
with all primary constraints. We have eight first class constraints for the 
seven pairs of canonical variables remaining after reduction. This seems 
to give the unphysical result that there are a negative number of degrees of 
freedom as in the $2D$ gravity models considered in \cite{Martinec}. However,
when counting degrees of freedom, we must only include the independent 
constraints. For the secondary constraints, the following relationship holds 

\begin{equation}
\label{19}\chi_1^{11}-\frac{2\Pi_0^{01}}{\Pi_0^{00}} \chi_0^{01} -
\frac 1{\Pi_0^{00} \Pi_0^{00} }\left( \Pi_0^{01}\Pi_0^{01}-1\right) 
\chi_1^{01} = 0. 
\end{equation}

This reduces the number of independent constraints to seven, which implies 
there are zero degrees of freedom, as expected. 

The presence of five primary constraints indicates that the group of gauge 
transformations has five parameters. Using the approach of Castellani  
\cite{Castellani1982}, we will recover the gauge transformation of all the 
initial fields.

The generator $G$ of the gauge transformation is found by first setting 
$G^b=C_P^b$ for the primary constraints that do not produce secondary first 
class constraints (i.e. $\left\{ C_P^b,H\right\} =0$) and 
$G_{\left(1\right) }^a=C_P^a$ for primary constraints that produce 
secondary constraints (i.e $\left\{ C_P^a,H\right\} \neq 0$). In this case
$C_P^b\in\left( \pi _{11},\Pi_0^{11}-\frac 1{\Pi_0^{00}} 
\left( \Pi_0^{01}\Pi_0^{01}-1\right) \right)$ and
$C_P^a\in\left( \Pi_1^{11} ,\Pi_1^{01},\Pi_1^{00}\right)$. 
We then introduce $G_{\left( 0\right) }^a\left( x\right)
=-\left\{ C_P^a,H\right\} \left( x\right) +\int dy\ \alpha _c^a\left(
x,y\right) C_P^c\left( y\right) $ where the functions $\alpha _c^a\left(
x,y\right) $ are found by requiring that $\left\{ G_{\left( 0\right)
}^a,H_c\right\} =0$. (In our model, not all of the functions 
$\left\{ C_P^a,H\right\} \left( x\right)$ are independent, which is a situation
distinct from what was considered in \cite {Castellani1982}. We do however 
construct a generator of a gauge transformation which leaves the action 
invariant even off shell.) The full generator of gauge transformations is then 
given by 
$$
G\left( \varepsilon ^b;\varepsilon ^a,\dot \varepsilon ^a\right) =\int
dx\left( \varepsilon ^b\left( x\right) G^b\left( x\right) +\varepsilon
^a\left( x\right) G_{\left( 0\right) }^a\left( x\right) +\dot \varepsilon
^a\left( x\right) G_{\left( 1\right) }^a\left( x\right) \right). 
$$ 
In our case this leads to the following expression
$$
G\left( \varepsilon \right) =\int dx\left[ \varepsilon ^{11}\pi
_{11}+\varepsilon _{11}\left( \Pi_0^{11}-\frac 1{\Pi_0^{00}} \left( \Pi_0^{01}
\Pi_0^{01}-1\right) \right)\right.  
$$
\begin{equation}
\label{21}\left.+\varepsilon \left( -\chi_1^{01} -\xi_{00}^1
\Pi _1^{00}+\xi_{11}^1\Pi_1^{11}\right) +\dot \varepsilon \Pi_1^{01} \right. 
\end{equation}
$$
\left.+\varepsilon _1\left( -\chi_1^{11}-2\xi_{01}^1\Pi_1^{11} 
-2\xi_{00}^1 \Pi_1^{01}\right) 
+ \dot \varepsilon _1\Pi_1^{11}+
\varepsilon ^1\left( -\chi _1^{00}+2\xi_1^{11}\Pi_1^{01} +2\xi_{01}^1 
\Pi _1^{00}\right) +\dot \varepsilon ^1\Pi _1^{00}\right].
$$

The PB of the generators (\ref{21}) has an algebra similar to what appears in
\cite {KKM1} when $h^{\alpha \beta}$ is treated as an independent field

\begin{equation}
\label{17}\left\{ G\left( \varepsilon \right) ,G\left( \eta \right) \right\}
=G\left( \tau ^c=C^{cab}\varepsilon ^a\eta ^b)\right) 
\end{equation}

where $\varepsilon ^a=\left( \varepsilon ^1(\varepsilon ),\varepsilon
^2(\varepsilon _1),\varepsilon ^3(\varepsilon ^1),\varepsilon ^4(\varepsilon
_{11}),\varepsilon ^5(\varepsilon ^{11})\right) $ and the only non-zero
structure constants $C^{cab}$ are $C^{132}=2=-C^{123},$ $%
C^{212}=1=-C^{221},C^{331}=1=-C^{313}$. This  reflects the structure of 
algebra of PB among all first class constraints.

Using (\ref{21}) we can determing the gauge transformation of the seven pairs 
of phase space variables remaining after the Dirac reduction by using the 
equation
$\delta field=\left\{ field,G\left( \varepsilon\right) \right\} $. 

The transformations of the fields $\Pi_1^{\alpha \beta} $ and $\xi_{\alpha 
\beta}^1 $ are the same as in \cite {KKM1} where $h^{\alpha \beta}$ is treated
as an independent field 

\begin{equation}
\label{23}\delta \Pi_{1}^{01} =\varepsilon _1\Pi_{1}^{11}-\varepsilon ^1
\Pi _{1}^{00},
\delta \Pi_{1}^{00}=\varepsilon \Pi _{1}^{00}+2\varepsilon _1\Pi_{1}^{01} ,
\delta \Pi_{1}^{11}=-\varepsilon \Pi_{1}^{11}-2\varepsilon ^1\Pi_{1}^{01}, 
\end{equation}

$$
\delta \xi_{01}^1 =\frac 12 \dot \varepsilon +\varepsilon ^1 \xi_{11}^1-
\varepsilon_1\xi _{00}^1,\delta \xi _{00}^1=\dot \varepsilon ^1-
\varepsilon \xi _{00}^1+2\varepsilon^1\xi_{01}^1,
$$
\begin{equation}
\label{25}\delta \xi_{11}^1=\dot \varepsilon _1+\varepsilon \xi_{11}^1-
2\varepsilon_1\xi_{01}^1. 
\end{equation}

There are slightly modified transformations for $\Pi_0^{\alpha \beta} $

$$
\delta \Pi_0^{00} =\varepsilon \Pi_0^{00} +2\varepsilon _1\Pi_0^{01},
\delta \Pi_0^{01}=\varepsilon _1\left( \frac 1{\Pi_0^{00}} \left( \Pi_0^{01}
\Pi_0^{01}-1\right) \right)-\varepsilon ^1\Pi_0^{00} ,
$$
\begin{equation}
\label{24}\delta \Pi_0^{11}=-\varepsilon \left( \frac 1{\Pi_0^{00}} \left(
\Pi_0^{01}\Pi_0^{01}-1\right) \right) -2\varepsilon ^1\Pi_0^{01}, 
\end{equation}

while those for $\xi_{\alpha \beta}^0 $ are quite distinct

\begin{equation}
\label{27}\delta \xi _{11}^0=\varepsilon _{11}, 
\end{equation}

\begin{equation}
\label{28}\delta \xi_{00}^0 =-\varepsilon _{,1}^1-\varepsilon \xi_{00}^0
+2\varepsilon ^1 \xi_{01}^0 +\frac {1}{\Pi_0^{00} \Pi_0^{00}}\left( \Pi_0^{01}
\Pi_0^{01}-1\right) \left(
\varepsilon _{11}+\varepsilon _{1,1} +2\varepsilon _1\xi _{01}^0-\varepsilon
\xi _{11}^0\right), 
\end{equation}

\begin{equation}
\label{29}\delta \xi_{01}^0 =-\frac 12\varepsilon _{,1}+\varepsilon^1 
\xi_{11}^0-\varepsilon _1 \xi_{00}^0 -\frac {1}{\Pi_0^{00}}{\Pi_0^{01}}
\left( \varepsilon _{11}+\varepsilon _{1,1} +2\varepsilon _1\xi _{01}^0
-\varepsilon \xi _{11}^0\right). 
\end{equation}

There is also a new pair of transformations

\begin{equation}
\label{22}\delta \pi _{11}=0,\delta g^{11}=\varepsilon ^{11}, 
\end{equation}

which reflects the disappearence, noted above, of $g^{11}$ from $H_c$ once 
the Dirac reduction has been performed.
We obtain the transformation of $h^{\alpha \beta}$ by using the first two 
of equations (\ref{24}) and strong equalities (\ref{9},\ref{12})

\begin{equation}
\label{30}\delta h^{00}=\varepsilon h^{00}+2\varepsilon _1h^{01},\delta
h^{01}=\varepsilon _1h^{11}-\varepsilon ^1h^{00}.
\end{equation}

Note, that the third equation of (\ref{24}) gives 
$\delta \Pi _0^{11}=-\varepsilon h^{11}-2\varepsilon ^1h^{01}$ 
but we cannot use this to find $\delta h^{11}$ as $\Pi_0^{11} \approx h^{11}$ 
only weakly. To determine the variation $\delta h^{11}$ we have to use 
(\ref{11}) or alternatively (\ref{13}) and then use (\ref{30}) to obtain

\begin{equation}
\label{32}\delta h^{11}=-\varepsilon h^{11}-2\varepsilon ^1h^{01}. 
\end{equation}

The transformation of $h^{\alpha \beta}$ we have found is the same as in 
\cite{KKM1} where $h^{\alpha \beta}$ is treated as being independent. 

Finally, the transformations for $g^{00}$ and $g^{01}$ can be found using the
strong equalities (\ref{9},\ref{10}) and the transformations 
(\ref{22},\ref{30})

\begin{equation}
\label{37}\delta g^{00}=\varepsilon ^{11}\frac{g^{00}}{g^{11}}+\varepsilon
2g^{00}+\varepsilon ^1\frac{2g^{00}g^{01}}{g^{11}}+\varepsilon _12g^{01}, 
\end{equation}

\begin{equation}
\label{38}\delta g^{01}=\varepsilon ^{11}\frac{g^{01}}{g^{11}}+\varepsilon
g^{01}+\varepsilon ^1\frac{2g^{01}g^{01}}{g^{11}}+\varepsilon
_1g^{11}-\varepsilon ^1g^{00}. 
\end{equation}

Equations (\ref{25},\ref{27}-\ref{22},\ref{37},\ref{38}) 
give the transformations 
of all fields in the original action (\ref{LEH}). (In (\ref{28},\ref{29}) we 
have had 
to express $\Pi_0^{00}$ and $\Pi_0^{01}$ in terms of $h^{\alpha \beta} 
\left(g^{\alpha \beta}\right)$ using equalities (\ref{9},\ref{12}).) To check 
the invariance of the Lagrangian we can use (\ref{30},\ref{32}), as 
$g^{\alpha \beta}$ enters 
$\tilde L_2$ only through $h^{\alpha \beta}$. In this case the fifth parameter 
of the gauge 
transformations $\varepsilon^{11}$ becomes ``hidden'' (as it enters only 
in transformations of $g^{\alpha \beta}$). However, as we have five primary 
constraints, there should be five parameters. Without this parameter 
it is impossible to obtain the transformations of all the components of 
$g^{\alpha \beta}$. 
All these peculiarities make the $2D$ action a very interesting model from 
point of view of constraint dynamics.

The variation of $\tilde L_2$ in (\ref{LEH}) gives 

\begin{equation}
\label{L} \delta \tilde L_2 = 2 \left[ \left(
h^{11}-\frac {h^{01}h^{01}}{h^{00}}
\right) 
\left(
\varepsilon _{11}+\varepsilon _{1,1} +2\varepsilon _1\xi _{01}^0-\varepsilon
\xi _{11}^0\right)
\right]_{,0}
\end{equation}

and the action is invariant provided this total derivative can be neglected.

It is not clear why the gauge transfromation that leaves $\tilde L_2$ invariant
in the $h$ formulation only leaves $\tilde L_2$ up to a total time derivative 
in the $g$ formulation. Possibly, the linear dependence of secondary 
constraints in (\ref {19}) has to be taken into account when applying the 
Castellani procedure. 

One can also find the transformation of the affine connection by using 
(\ref{ICT2}). \\

{\bf CONCLUSION}\\

We have demonstrated that the canonical procedure applied 
to the EH action in $2D$ with $g^{\alpha \beta}$ being independent leads to 
the same canonical structure as in the approach where the metric density 
$h^{\alpha \beta}$ is an independent field. It also leads to a gauge 
invariance which is different 
from a diffeomorphism. This is despite having a different number of constraints
in the two formulations.

Based on these results in $2D$ and some preliminary investigations of the 
higher dimensional EH action, we are tempted to conclude that these canonical
properties should be preserved for gravity in all dimensions.
We think that it is highly unlikely
that the theory (first order formulation of EH), which behaves as a local field
theory in $2D$, is not a local field theory when $d>2$. Beyond two 
dimensions, straightforward application of the Dirac approach is complicated 
(even with the simplification of using $\xi_{\alpha \beta}^\lambda$ in place 
of $\Gamma_{\alpha \beta}^\lambda$ ) by the 
fact that the Hamiltonian is no longer a linear combination of secondary 
constraints as in (\ref {14}) and furthermore at least tertiary constraints 
appear. 

If $d>2$, all secondary constraints coresponding to the momenta $\Pi _m^{nk}$ 
and 
$\Pi _{n}^{0k}$ (\ref{Md}) (except the momentum $\Pi _k^{0k}$) constitute a 
special set of second class constraints which can be eliminated. Thus, 
if $d>2$, 
there are $d$ primary and $d$ secondary constraints and they produce $d$ 
tertiary constraints. Although this work is in progress, we can make the 
argument that if the $d$ tertiary constraints are also first
class and the Dirac procedure does not lead to any further constraints, than 
we have $3d$ first class constraints. The fields left after 
using the second class constraints to eliminate some variables through the 
Dirac reduction are the $\frac {1}{2}d\left( d+1 \right)$ components of the 
symmetric second rank tensor ($\xi_{\alpha \beta}^0 $) 
plus the $\left(d-1\right) $ components of $\xi _{0k}^0$ plus the one 
component 
$\xi _{0l}^l$; this gives the total number of variables after the Dirac 
reduction to be $\frac {1}{2}d\left( d+3\right)$. Subtracting the number of 
first class constraints ($3d$) leaves
us with $\frac 12d\left( d-3\right) $ degrees of freedom. This is the number 
of degrees of freedom present in a spin two gauge field in any dimension.
Of course, this scenario with $3d$ first class constraints is not the only 
possibility that gives a
correct expression for this number of degrees of freedom, but it does 
illustrate that the presence of tertiary constraints does not contradict
having the anticipated number of degrees of freedom.

The ADM \cite{ADM1,ADM2} analysis of first order formulation of the EH action
ends with secondary constraints where we, using the Dirac reduction, have at 
least tertiary constraints. The 
difference between the two approaches is based on the fact that in the ADM
procedure not only second class, but also first class constraints were solved 
during the preliminary Lagrangian reduction of the action. This is because all 
time independent equations of motion are used to eliminate some variables 
without distinguishing whether these equations correspond to first class or 
second class constraints. (See a very clear 
exposition of this procedure in Appendix I of the review article 
\cite{Fad1982}.) Solutions of the
30 constraint equations (the number of field equations without time 
derivatives of fields in $4D$) after substitution back into the original 
Lagrangian leads to a disappearance of 34 variables (a clear indication that 
four first class constraints have been solved) and the reduced action hence 
has only 
16 variables (see Eq.(4.1) of \cite{ADM2}). In our approach in which only 
the second class constraints are eliminated, after the Dirac reduction 
24 variables are left. A more detailed discussion of this point will be 
published elsewhere.

An important question for the higher dimensional EH action is whether it is 
possible to preserve all canonical properties present in the $2D$ case by 
following the standard Dirac procedure without any {\it a priori}
assumptions and restrictions, and to determine what gauge transformation 
this procedure produces. 
(It is possible that a diffeomorphism is not {\it the only} 
symmetry of $d$ dimensional EH action; we have seen that in fact an 
alternate symmetry occurs in $2D$.) If the final
algebraic structure of PB of first class constraints is local, then a viable 
approach to quantizing higher dimensional gravity may exist. \\

{\bf ACKNOWLEDGMENTS}\\

Authors would like to thank members of the theory group seminar here, at
Western, for comments and suggestions and especially A.Buchel and E.V.Gorbar.

N.K. and S.V.K. would like to thank S.B.Gryb for continuing interest to this
work and stimulating discussions.

D.G.C.McKeon would like to thank NSERC for financial support and R. and D.
MacKenzie for useful suggestions.

\end{document}